# Disorder Induced Superconductivity in $TiSe_{1.2}S_{0.8}$

M. Singh[1], P. Saha[1], A. Chahar[2], B. Birajdar[2], D.K. Shukla[3], S. Patnaik[1,*]

[1]*School of Physical Sciences, Jawaharlal Nehru University, New Delhi 110067, India*

[2]*Special centre for Nano Sciences, Jawaharlal Nehru University, New Delhi 110067, India*

[3]*UGC-DAE Consortium for Scientific Research, Indore, Madhya Pradesh 452017, India*

*Corresponding author: spatnaik@jnu.ac.in*

## Abstract

Disorder can be utilized as an effective parameter to probe the interplay between two long range orders such as superconductivity and charge density wave. In the present work, we report on the experimental evidence for filamentary superconductivity in polycrystalline $TiSe_{1.2}S_{0.8}$ with superconducting transition $T_c \sim 7K$. This is validated from magnetization and magneto-transport measurements. Strain induced dislocations, substitutional defects, and randomly distributed Ti ions (with local moments) are considered as possible sources of disorder. A detailed analysis of the temperature dependent resistivity evaluates the degree of disorder and the consequent localization effects. The findings are in striking contrast to the fact that superconductivity has not been reported in single crystals of $TiSe_{2-x}S_x$ system. It is established that disorder serves as a stabilizing factor for the superconducting phase due to in-commensuration of the charge density wave.

# Introduction

The emergence of superconductivity in cuprates, pnictides and dichalcogenides is generally ascribed to the interplays of the underlying charge or magnetic ordering [1-3]. It is understood that the associated phenomena such as the charge density wave (CDW), spin density wave (SDW), or the nematic order (NO) need to be suppressed towards attaining the optimal superconductivity. The origin of the condensate pairing mechanism is derived from such competing long- range orders. However, there is increasing evidence that such ordered states and lattice disorder, instead of competing, suppressing, or co-existing, may actually aid the attainment of superconducting state [6-10]. In this paper we report on the experimental manifestation of filamentary superconductivity in TiSeS, that is evidenced to be driven by disorder and charge ordering. This result has wider ramification since the layered transition metal dichalcogenides (TMDs) bear similarity in the observed phase diagram to those of cuprates and other unconventional superconductors [11-15].

It is well established that disorder plays a crucial role in shaping the electronic properties of all correlated systems. Disorder may lead to Anderson localization causing distortions in the long-range order [6-10,16]. The effects of disorder on the competition between superconducting and CDW phases have been well documented [9,16]. Recent studies on disordered systems shows that incommensuration and fragmentation caused by disorder play a key role in driving superconductivity in quasi 2D systems [17-23]. Theoretical studies have implied that the presence of disorder can actually aid the onset of filamentary superconductivity where there is strong interaction between SC and CDW orderings [7,21]. It is summarized that disorder can serve as a tuning parameter between various electronic phases that result from the intertwining of disorder, CDW, and superconductivity [7,21,22]. The experimental realization of such a premise is mostly pending.

Among the TMDs, $TiSe_{2-x}S_x$ system has attracted substantial current interest [26-28,30]. While some studies have focused on the CDW state using tunnelling spectroscopy, others have highlighted on the thermoelectric properties, metal-insulator transition, and phonon-drag effects [24-36]. It is established that sulphur doping in place of selenium in $TiSe_2$ tends to suppress the CDW phase. The question therefore is whether disorder induced by substitution can also play a role in the suppression of CDW in $TiSe_{2-x}S_x$ that can lead to formation of superconducting state. Previous experimental reports as well as first-principle calculations have not provided much evidence for a possible superconducting order in TiSe$_{2-x}$S$_x$ (0<x<2) [27,28,30].

In the present study, the possibility of filamentary superconductivity in off-stoichiometric TiSeS is articulated. Magnetization data clearly shows a diamagnetic transition at 7K with clear hysteresis in ZFC and FC data. Resistivity data shows a broadened CDW transition near 60K. Magnetoresistance shows shifting of resistive transition towards the low temperature side on increasing magnetic field. A detailed analysis of the resistivity suggests that the electron-electron interactions and disorder lead to localization effects at low temperature. The possible origin of the reported filamentary superconductivity is explained on the basis of disorder and CDW, where the enhanced disorder tends to disrupt the CDW coherence but leads to metastable condensation.

**Experimental Techniques**

Polycrystalline TiSeS sample was prepared by solid state reaction method. Mixture of Titanium (Ti), Selenium (Se) and Sulphur(S) were taken in the nominal ratio and ground well in agate mortar-pestle to homogenise it. The mixture was then vacuum sealed in quartz tube and heated in a programmable furnace in two steps. Initially calcined at 650 °C for 4 hours and after that at 700°C for 24 hours followed by natural cooling. Shiny black polycrystalline sample was obtained. The post synthesis sintering of the prepared pellets was done to ensure grain growth and connectivity at 700°C for 24 Hours. Phase identification was done using Rigaku Miniflex-600 powder X-ray diffractometer. High resolution transmission electron microscopy (HRTEM) measurements were performed using a JEOL (JEM-2100F) microscope. A Zeiss EVO40 SEM analyser and a Bruker AXS microanalyzer were used for energy dispersive X-ray spectroscopy (EDX) and scanning electron microscopy (SEM) imaging, respectively. Magnetic measurements were done using a Quantum Design SQUID MPMS3. Transport measurements were performed in Quantum Design 14 Tesla PPMS.

**Results and Discussion**

Powder X-ray diffraction pattern of synthesized TiSeS is shown in Figure 1. The peaks are indexed with the corresponding Miller index *(hkl)* values. The patterns confirm with the $P_{-3m1}$ (164) space group having a trigonal lattice. As depicted in the inset, the atomic arrangement follows octahedral coordination. Reitveld refinement using the Fullprof software shows excellent matching with the calculated data. Earlier reports [26-28] confirm that the atomic coordination consists of S/Ti/Se in three separate layers. But in reality, in a solid solution, the possibility of occupation of Se and S atoms is equiprobable at each chalcogen site. This can lead to intrinsic disorder in the compound. The unit cell parameters deduced from refinement are estimated as a = b = 3.475 Å and c = 5.90 Å. This value is in-between the values of the corresponding parent compounds i.e. $TiS_2$ and $TiSe_2$. Inset of Figure 1 also includes data from SEM-EDX measurements for the determination of elemental composition in the as grown sample. The elemental ratio is found to be 1:1.2:0.8 for Ti:Se:S (normalized with respect to Ti atoms). Such characteristics is not un-common in TMDs because of the volatile nature of chalcogens. The EDX result validates that the compound is off-stoichiometric. Furthermore, the unit cell parameter of as grown $TiSe_{1.2}S_{0.8}$ matches very well with the reported cell parameters for x= 0.8 in $TiSe_{2-x}S_x$ [25]. The interesting fact is that such deviations from nominal composition can give rise to a variety of novel and emergent phenomena in TMDs. Studies on $TiSe_2$ have shown that slight change in selenium vacancies can change the transport behaviour from semiconducting to metallic [37]. $TiSe_2$ with the least Se vacancies itself is an indirect band semiconductor [37,38]. The resistivity anomaly associated with CDW transition also becomes prominent with the optimum amount of selenium vacancies [37].

In order to study the microstructural properties, the high-resolution transmission electron microscopy (TEM) measurements were conducted. For the aforementioned studies, TEM samples were prepared by ultrasonication. Figure 2(a) shows the bright field image with the smallest selected area aperture. Obtained Selective Area Electron Diffraction (SAED) pattern and HRTEM images are displayed in Figure 2(b) and Figure 2(c), respectively. SAED pattern is indexed with the corresponding (*hkl*) values. It shows a very good resemblance with the powder XRD data. It is verified by comparing the values of the interplanar distances (*d*) from the SAED pattern rings and those calculated from XRD. High resolution images show lattice fringes extending over distances greater than 35 nm. This indicates the grain size to be at least about 35 nm. This matches very well with the grain size determined using XRD. The HRTEM images shown in figure 2(c) depicts the crystallites of different orientation (indicated by blue arrows) separated by boundaries. Intragrain dislocations in the form of wrinkles were visible (dark blue circle) in figure 2(c). The strain brought on by the S atom substitution's size mismatch is the possible source of these dislocations. Therefore, the TEM studies have clearly brought forth the intra-grain disorder that possibly can affect the incommensuration of the CDW domains.

**Magnetization**

The magnetization curves under zero field cooled warming (ZFCW) and field cooled warming (FCW) were obtained at 20 Oe externally applied magnetic field. These are displayed in Figure 3(a). The magnetic field was applied perpendicular to sample plane. At T~ 7 K, the ZFCW data displays a diamagnetic change, which is indicative of a superconducting transition. Additionally, it is evident from the FCW data that they are separated from the ZFCW data, confirming that superconductivity is the cause of the diamagnetic transition at around 7 K. The superconducting volume fraction is very small because the order of moment is very small in ZFCW data. Another transition close to 4 K is also observed. It is possible that the 4 K transition corresponds to intragranular and intergranular crossover for the diamagnetic state. Here the grains are decoupled in the temperature range $T_{inter}<T<T_c$. Superconductivity is attained first at $T_c$ ~7 K (intragranular) followed by the temperature $T_{inter}$ ~4 K (intergraular). A similar situation has also been noted in granular iron-based superconductors and cuprates [39-41]. Such granular superconductivity is supported by the grain size determined by the TEM patterns.

To consolidate the evidence for superconductivity, magnetization measurements were carried out at different external fields with H applied parallel to sample plane for H =20 Oe, 100 Oe, 500 Oe). This is shown in Figure 3(b), 3(c) & inset of 3(c) respectively. It is evident that when the field is increased, the $T_c$ moves to the lower temperature side, which is indicative of suppression of superconducting state in the presence of external magnetic field. We note that at 500 Oe the moment becomes positive for both ZFCW and FCW. This is due to dominance of paramagnetic background of the sample. Inset of figure 3(a) and 3(b) shows the MH loop taken at 1.8 K in two different sample geometries, H⊥ and H∥ to sample plane respectively. This shows the type-II superconductor like irreversible M-H loop. Temperature evolution of the magnetization is shown in Figure 4(a) which consists of M-H loop at 1.8 K, 4 K and 10 K respectively with field applied perpendicular to the plane. It is evident that there is no superconductor-like hysteresis at 10 K, which is the normal state (above $T_c$). We further note that the moment at the maximum field 500 Oe decreases as the temperature is increased. This is suggestive of a shrinking of filamentary superconducting state in a paramagnetic background. The important parameter which can be derived from the M-H loop are the critical current density ($J_c$) and lower critical field ($H_{c1}$). To estimate $H_{c1}$, M-H data at different temperatures were taken upto 100 Oe. The method of deviation from Meissner line was utilized to determine the lower critical field. The Meissner line is the linear behavior of M-H curve in low field region. The deviation ($\Delta M$) is calculated as $\Delta M = M - \frac{dM}{dH}H$ where $dM/dH$ is the slope of M-H curve in low field region and H is the applied field. The $\Delta M$ curves at various temperatures are shown in inset of Figure 4(b). The criterion of determining the $H_{c1}$ was chosen to be $\Delta M = 1 \times 10^{-6}$ emu [43]. Figure 4(b) illustrates variation of $H_{c1}$ with temperature. The temperature dependence of $H_{c1}$ reflects parabolic characteristics as expected from Ginzburg-Landau formalism. With the GL approximation, the value of $H_{c1}(0)$ estimated from the best fit comes out to be $\approx$ 53 Oe. This implies a zero-temperature penetration depth to be $\approx$ 250 nm as derived from the formula. A rough estimate of the critical current density $J_c$ is obtained by the Bean model formula $J_c = \frac{20\Delta M}{w(1-\frac{w}{3l})}$. Here $l$ and $w$ are the length and width of the sample respectively with $w < l$ [44]. The difference in magnetization ($\Delta M$) is deduced from M-H loop using the relation $M(H_-)-M(H_+)$. The magnetization values $M(H_-)$ and $M(H_+)$ are the moment (in emu/cc) during the field decrease and increase cycles respectively [44]. The field dependence of $J_c$ at 1.8 K and 4 K are shown in inset of Figure 4(a). The maximum obtained value of $J_c$ at 1.8 K is estimated to be $14.8 \times 10^4$ Amp/m$^2$. This maximum is very sensitive to the grain size. It usually shifts towards the lower field side and high $J_c$ values for smaller grains as reported in Iron based superconductors and High $T_c$ superconductors (HTS) [42]. Further the $J_c$ decreases in gradual manner rather than rapidly with field, which indicates towards reasonable vortex pinning in the sample due to granularity. Both intragrain pinning and inter-grain pinning are responsible for the pinning process at 1.8 K. However, the inter-grain pinning mostly plays a dominant role because of the filamentary character that lacks the appropriate grain connectivity between the superconducting zones.

**Transport Properties**

The electrical resistivity measurements were performed using linear four probe method. The low temperature resistivity is shown in Figure 5(a). It is seen that resistivity increases on lowering the temperature instead of attaining zero resistance as it should be for bulk superconductivity. Such localization behavior in resistivity at low temperature is in resemblance with earlier reports on TiSeS [27]. Only a slight deviation towards the lower resistance is observed near 4 K. Similar deviation was also observed in magnetization data ZFCW. Since the magnetization data concluded a very small superconducting volume fraction, hence due to weak grain connectivity between superconducting regions no observable zero-resistance state could be attained. Earlier reports on filamentary superconductivity in $ZrSe_2$ and $WO_{2.90}$ [45,46], have reported similar transition in magnetization but not in resistivity due to the fact that superconductivity is localized in small regions and percolative current across the sample is not attained. It hints towards the fact that the observed superconductivity in $TiSe_{1.2}S_{0.8}$ is of filamentary nature. When discussing the origin of superconductivity in filamentary superconductors, two potential scenarios have been explored. In the case of $ZrSe_2$, the Zr vacancies lead to partially filled bands that offer a superconducting platform. On the other hand, in tungsten oxide systems, charge carrier 1D stripes are created as a result of bipolaron production and their clustering across shear planes leads to superconductivity. Some other reports on pnictide superconductors [47] have also shown the possibility of filamentary superconductivity in doped pnictides, where suppression of anti-ferromagnetic fluctuations provide the glue to form Cooper pairs.

Following the interpretations from the magnetization data, resistivity analysis can be done in more detail. The complete resistivity curve (Figure 5(a)) can be divided into four temperature ranges; (i) below 7 K where superconducting transition is observed, (ii) localization behavior in temperature range (8 K-40 K), (iii) 40 K-240 K where a hump-like structure is observed, and (iv) 240 K-300 K where a linear variation of resistivity with temperature is observed. In region (ii) resistivity shows the minima at around 35K that increases as the temperature is lowered. This kind of behavior is indicative of electron-electron interactions and localization effects [27,48-50]. Such localizations can originate from various types of interactions, such as dilute magnetic impurities that can give rise to Kondo insulating phase [48], Anderson localization owing to disordered systems [49], strong electron-electron interactions that can give rise to Mott insulators and Weak localization (WL) [50]. The question is what kind of competing interactions may lead to the upturn in resistivity in $TiSe_{2-x}S_x$. In previous reports on thin films and alloys [51-53], the possibility of quantum interference effects (QIE) such as electron-electron (e-e) interactions and WL are discussed as the origin of such localization behavior. Taking elastic e-e interactions, the data below resistivity minimum were fitted with the following equation;

$$\rho(T) = \rho_0 + A\sqrt{T} + BT^n \tag{1}$$

Here the first term is the residual resistivity, the second term relates to electron-electron interactions in weak localization limit in disordered materials [56]. The third term is a generalized temperature dependent term where $n$ can take value according to the interactions such as magnetic, electron-phonon, or inter-band scattering. For $n = 2$ the equation fit perfectly to the experimental data (shown in inset (i) of Figure 5(a)). Obtained value of parameters are estimated to be $\rho_0 = 20.72882$ mΩ-cm, $A = -0.798$ mΩ-cm K$^{-1/2}$, $B=9.771\times10^{-4}$ Ω-cm K$^{-2}$. Charge density fluctuations give rise to $T^2$ dependence to resistivity at low temperatures for pure paramagnetic transition-metal based compounds [55]. Scattering of conduction electrons with localized moments also leads to a $T^2$ dependency [55,56]. This assumption complements our results well. This is because in TiSe$_{2-x}$S$_x$, Ti local moments are present due to Ti vacancies or Ti occupying interstitial sites [27,31]. We note that a higher growth temperature (700°C) increases the possibility of Ti intercalation [31]. A high order of moment (10$^{-2}$ emu/mol Oe) was measured in the sample in 1 Tesla external applied field at 2 K. Such high moment can arise because the Ti atoms that are occupying the interstitial sites (with 3d$^1$ or 3d$^2$ configuration), can rise to local moments. However, those bonded to chalcogen atoms will have non-magnetic 3d$^0$ configuration [57]. Hence, in TiSe$_{1.2}$S$_{0.8}$, magnetic contribution arising from localized moments as well as charge density fluctuations justify the observed $T^2$ dependency.

A broader hump-shaped anomaly is observed in resistivity (Figure 5 (a)), which lies in the 3$^{rd}$ temperature range between 50 K-120 K. It can be attributed to CDW transition in the compound. The broadness of the hump in the data is suggestive of high intrinsic disorder in the compound. In the previous report on TiSeS (x = 1) single crystals, no CDW transition was observed [27]. With changing Se:S ratio transport properties of TiSe$_{2-x}$S$_x$ [29-31] show appearance of CDW phase. The CDW transition temperature in TiSe$_{1.2}$S$_{0.8}$ as obtained from the derivative with respect to temperature is found to be 60 K (inset of Figure 5(b)). The criteria used to define the CDW temperature is same as described in reference [59]. Further, CDW transitions are characterized by thermal hysteresis [60,61]. A resolvable hysteresis is also observed in the sample (Figure 5(b)). The presence of the CDW transition can be justified by the off-stoichiometry of the sample.

The linearity of resistivity in the range of 240 K to 300 K is shown in the insert (ii) of Figure 5(a). This linear trend can be attributed to the dominance of electron-phonon scattering at high temperature. The resistivity can be fitted above $T_{CDW}$ using the formula $\rho(T) = a'T+b'$, where $b'$ results from local CDW fluctuations and $a'$ is the factor related to electron-phonon scattering [16]. From the fitted data, and the respective values of $a'$ and $b'$ are estimated to be 0.0998 mΩ-cm K$^{-1}$ and 0.44209 mΩ-cm. For CDW materials, the constant $b'$ correlates to the degree of disorder where not all states are gapped out below $T_{CDW}$. So, in presence of disorder the value of $b'$ is a measure of the suppression of CDW phase and corresponding increase in the carrier density [16].

The data from 8 K up to 35 K (inset (i) of Figure 5(a)) were fitted using the electron-electron correlation having both WL as well as the quadratic ($T^2$) terms with an excellent accuracy ($R^2$= 0.99999). This extrapolation was used to determine the expected behaviour at low temperatures (below 8 K) in the presence of WL and $T^2$ terms. The low temperature resistivity data between 1.8 K to 15 K was then subjected to point by point subtraction of the corresponding extrapolated background data. As seen in Figure 6 (a), a clear drop in resistivity is observed below 7 K This unambiguously reconfirms the onset of superconductivity in TiSe$_{1.2}$S$_{0.8}$ in conjunction with the magnetization data.

Furthermore, the density of states (DOS) at Fermi energy $N(E_F)$ can be quantitatively understood by the coefficient of $\sqrt{T}$ term in the conductivity [52, 56]. For the temperatures up to ($T \leq T_{min}/2$), with $T_{min}$ being the temperature where minima in resistivity occurs (~ 40 K ), the electrical conductivity due to electron-electron interactions can be given by $\sigma(T) = \sigma(0) + C\sqrt{T}$. In the given temperature range, the fit using the aforementioned relation gives the value of $C = 1.9$ (Ω-cm K$^{1/2}$)$^{-1}$. The coefficient of $\sqrt{T}$ term ($C$) is related to the diffusion constant ($D$) by the following relation

$$C = \frac{1.3e^2}{3\sqrt{2}\pi^2 \hbar} \left[\frac{k_B}{\hbar D}\right]^{1/2} \qquad (2)$$

The Einstein relation for resistivity is given by $\rho = 1/ne^2 N(E_F) D$ [54], where $D$ is the diffusion constant evaluated from Eq. (2). Using the Einstein equation (for $\rho = \rho_0$), the calculated value of DOS at Fermi energy comes out to be ~ 10$^{45}$ (J.m$^3$)$^{-1}$. The obtained value is somewhat lower than values found in other TMDs [62]. This is due to the presence of CDW in the material that causes the reduction of DOS at Fermi level. A lower DOS is the result of enhanced inherent disorder. One more possible reason of the lower density of state may be the presence of the semiconducting gap at lower temperature that originates from the S substitution. It opens up the gap in the electron-hole bands and results in the localization of charge because of the presence of 3p orbitals of sulfur [30]. It is to be noted that the sulfur content in our sample is in good agreement with the sulfur concentration at which gap opening occurs (x ~ 0.8) [30].

In the inset (i) of Figure 6(a), magnetoresistance (MR) data are plotted at T = 2K with the field applied perpendicular to sample plane. The data depicts the % MR, calculated using the formula, % MR = [{R(H)-R(0)}/R(0)] ×100%. In the observed data, the MR initially increases with the applied magnetic field, then it gets saturated near 3000 Oe and remains nearly constant upto 6000 Oe. The observed behaviour is expected for the magnetoresistance curve for a typical type -II superconductor. The increasing resistance in a step-like manner at lower field shows that a superconducting phase is present, which gets suppressed after reaching the upper critical field. The increase in MR is more pronounced up to the applied field of 450 Oe, then it becomes more gradual beyond 450 Oe. The reason for change of slope is related to crossover from inter-grain to intra-grain suppression. The magnetoresistance curves up to 60 kOe magnetic field at various temperatures, 2K, 10K, and 50K, respectively are displayed in inset (ii) of Figure 6(a). At 2 K in higher magnetic field range the MR is dominated by localization effects, causing negative MR. The system might have had only a negative magnetoresistance (MR) if there had not been a superconducting phase. At 10 K (above $T_c$), the plot shows negative MR in complete field range. The presence of negative MR also confirms the presence of the local Ti moments. The Lorentz force contribution alone would produce a positive MR with a typical parabolic field dependency. The negative contribution comes from the interaction between localized moments and conduction electrons. In the presence of magnetic field, the reduction in the scatterings caused by the local moments gives rise to negative MR [56,57]. At 50 K the MR is negative with nearly linear behavior. It is because at higher temperatures, due to greater thermal fluctuations, the spin-flip scatterings caused by local moments become less effective. MR at low temperatures (1.8 K, 2 K, 2.2 K, 2.5 K) up to 1000 Oe magnetic field is shown in Figure 6(b). A clear shift in plateau-like structure is observed towards lower field side, with increase in temperature. It is reflective of decrease in upper critical field as $T_c$ is approached. Since the measured MR is up to 1000 Oe, the shift in inter-grain transition could be observed. The onset of the transition shifts from ~560 Oe for 1.8K to ~320 Oe for 2.5K. The observed shift again validates the existence of superconducting state. The magnetization and magnetoresistance measurement are therefore consistent with the existence of filamentary superconductivity in TiSe$_{1.2}$S$_{0.8}$. It is evident from the observed results that the disorder plays an important role in both transport and magnetic properties of TiSe$_{1.2}$S$_{0.8}$. The transport properties near CDW temperature are dominated by the disorder, leading to a less intense or a fragmented CDW phase. The low temperature localization behaviour is again a consequence of disorder enhanced scatterings. The influence of this disorder in inducing filamentary superconductivity in TiSe$_{1.2}$S$_{0.8}$ is the next question that is addressed below.

Several recent reports on varied systems have reported superconducting state being influenced by disorder [6,8,10,16,23]. The central thesis is that disorder leads to suppression of the CDW fluctuations and thereby leading to the enhancement in superconducting $T_c$. In TiSe$_{1.2}$S$_{0.8}$, the isoelectronic substitution of sulphur cause suppression of CDW temperature ($T_{CDW}$~100K) and broadening in CDW transition as well as reduction in CDW amplitude due to enhanced disorder. The question arises that why the superconductivity is filamentary in TiSe$_{1.2}$S$_{0.8}$? The origin can be understood by the fragmentation or incommensuration of a commensurate CDW (CCDW) phase due to disorder that causes the formation of domain walls [17-23]. This matrix of ICCDW which is embedded in matrix of CCDW leads to the formation of superconducting order parameter [6-8,17-23]. The observed spatial modulations of the Cooper pairs in earlier experiments have suggested the requirement of an inhomogeneous electronic state matrix to pin the superconducting condensate. Further, the phase fluctuations or the phonon modes of the ICCDW could lead to formation of a local condensate of Cooper pairs. Thus, the incommensurate CDW regions dictated by lattice disorder plays an important role in emergence of condensate formation. However, in the presence of a strong competition between the two order parameters i.e. SC and CDW, a metastable phase i.e. filamentary superconductivity, could exist only at the phase boundary [7,21]. Disorder present in the system, plays a key role and stabilizes the filamentary phase and possibly enhancing the $T_c$ to 7K which is one of the highest in TMDs. So, the enhanced disorder not only causes the occurrence of the superconducting order in the system but also possibly inhibits the formation of a long-range coherent bulk superconducting state.

**Conclusion**

In conclusion, we report experimental evidence for filamentary superconducting phase in polycrystalline TiSe$_{1.2}$S$_{0.8}$. Disorder in the form of strain induced wrinkle shaped intragrain dislocations were confirmed from in HRTEM images. A clear diamagnetic transition at 7K is observed followed by hysteresis in ZFC and FC curves. Resistivity data confirm localization effects. A resistive transition concurrent with diamagnetic transition is observed after background subtraction. Intra-grain dislocations caused by isoelectronic sulphur substitution and intercalated Ti local moments are included as possible sources of disorder. Presence of disorder not only causes the broadening in the CDW transition at intermediate temperature range but also results in the localization in low temperature range. The low density of state at Fermi level is assigned to charge localization effects. The observed superconductivity is assigned to the fragmentation of CDW phase driven by disorder. The filamentary nature is considered as a metastable phase which is stabilized by disorder. Thus, $TiSe_{1.2}S_{0.8}$ validates the existing theories that disorder can drive superconductivity in the presence of charge density wave with transition temperature multiple-times higher than most transition metal dichalcogenides.

## Acknowledgement

Mr. M. Singh acknowledges CSIR for providing CSIR-SRF fellowship. S. Patnaik thanks DST-NANOMISSION-CONCEPT, India - 10/2019(G)/6 for consumables and equipment grants. We thank UGC-DAE-CSR (Indore) for high resolution magnetization and resistivity measurements.

**Figure Captions:**

**Figure 1.** Refined XRD pattern of the powdered sample. Peaks are indexed with corresponding Miller indices (hkl). Inset shows the EDX spectra of the sample. Schematic crystal structure of TiSeS is also shown.

**Figure 2** (a) TEM Bright Field image with smallest SAED aperture inserted. The corresponding SAED pattern indexed with the (*hkl*) values, is shown in (b). (c) HRTEM image of the region marked in (a), showing grains of typical size 35-40 nm.

**Figure 3** (a) ZFC-FC curve at 20 Oe (H ⊥ Sample plane) showing diamagnetic transition at 7K. Inset shows the M-H loop at 1.8 K. (b) ZFC-FC curve at 20 Oe external field (H // Sample plane). Inset is the corresponding M-H loop. Figure (c) shows ZFC-FC curve at 100 Oe and its inset shows data for 500 Oe external field with field applied parallel to the sample plane. Lowering of $T_c$ with increasing external field is evident.

**Figure 4** (a) Temperature evolution of M-H loops (at 1.8 K, 4 K, 10 K). Inset is the critical current density obtained from Bean's Model at 1.8 K & 4 K plotted as the function of magnetic field (*H*). (b) $H_{c1}(T)$ values plotted as a function of temperature. Inset depicts the criterion used to determine $H_{c1}(T)$.

**Figure 5** (a) Resistivity curve as a function of temperature in the range (1.8 K to 300 K). Inset (i) shows the low temperature resistivity fitting. High temperature linear fit of resistivity (240 K-300 K) is shown in inset (ii). (b) Expanded CDW transition with thermal hysteresis. First derivative of resistivity in the inset shows $T_{CDW}$.

**Figure 6** (a) Low temperature resistivity data after subtracting the background (obtained from low temperature fitting of experimental data *[Δρ = ρ - ρ_{fitted}]*. It shows the onset of resistive transition at ~7 K. Inset (i) plots the % MR = *[R(H)-R(0)/R(0)] ×100*, in the low field range. Inset (ii) shows the % MR in the field range (up to 6 Tesla) at 2 K, 10 K and 50 K respectively. (b) plots the % MR at different low temperatures (1.8 K, 2 K, 2.2 K, 2.5 K) up to 1000 Oe magnetic field. Shift of the transition to lower field side with increasing temperature is seen.

**Fig 1.**

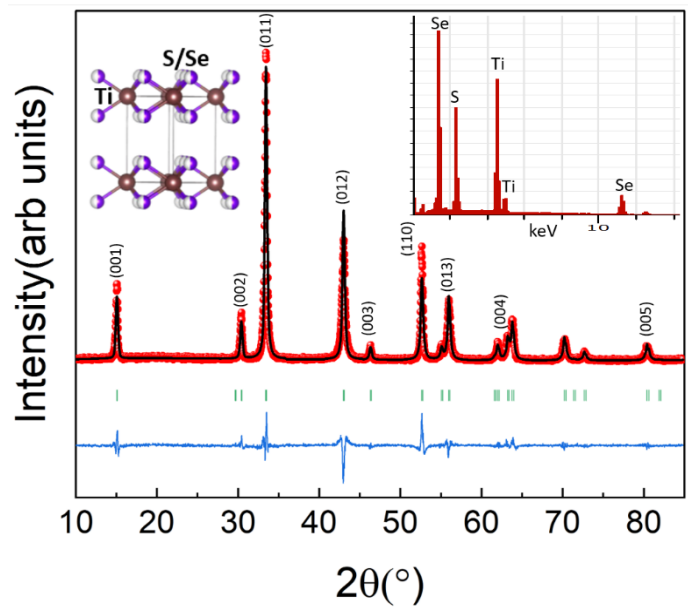

**Fig. 2**

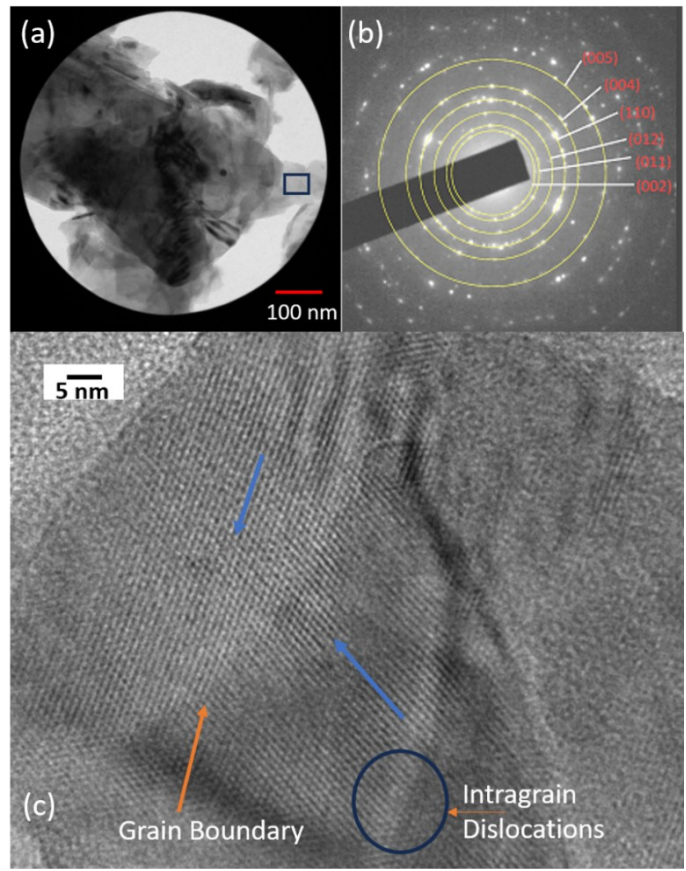

**Fig 3**

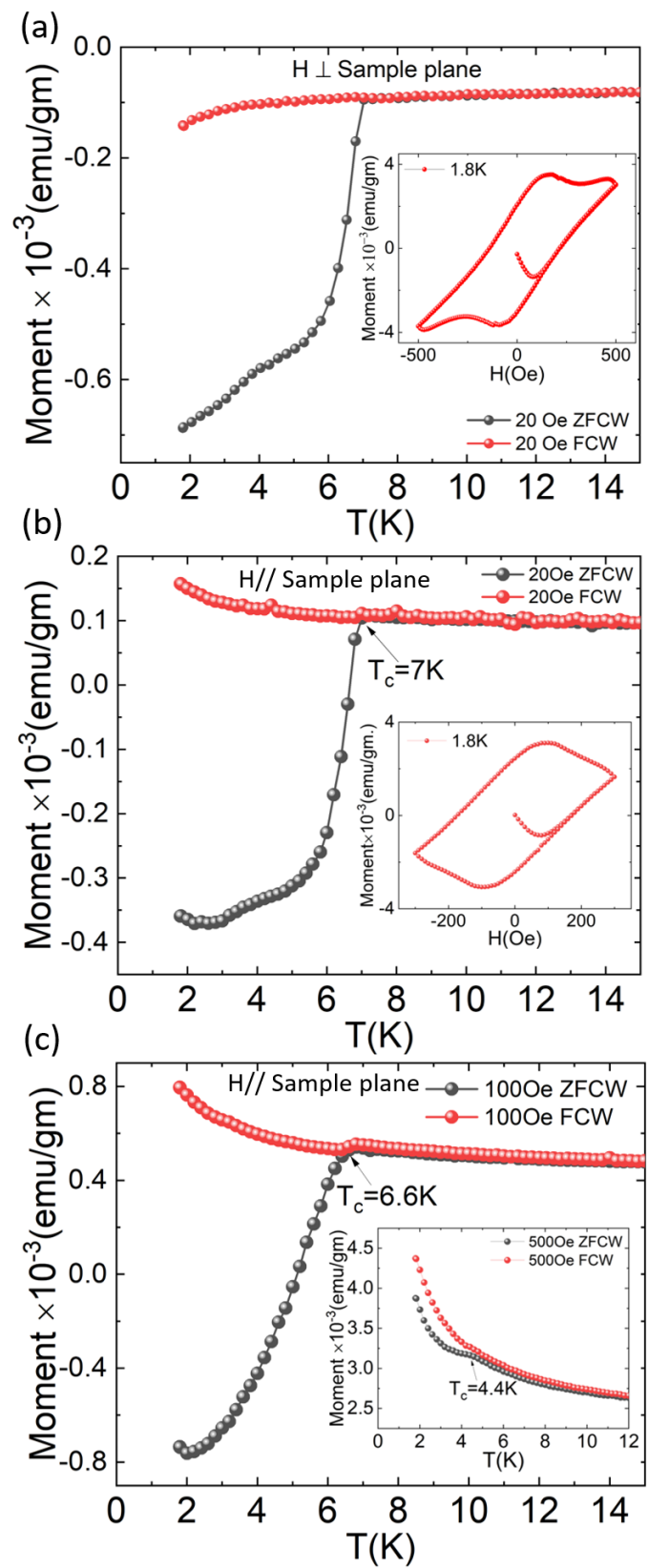

**Fig 4.**

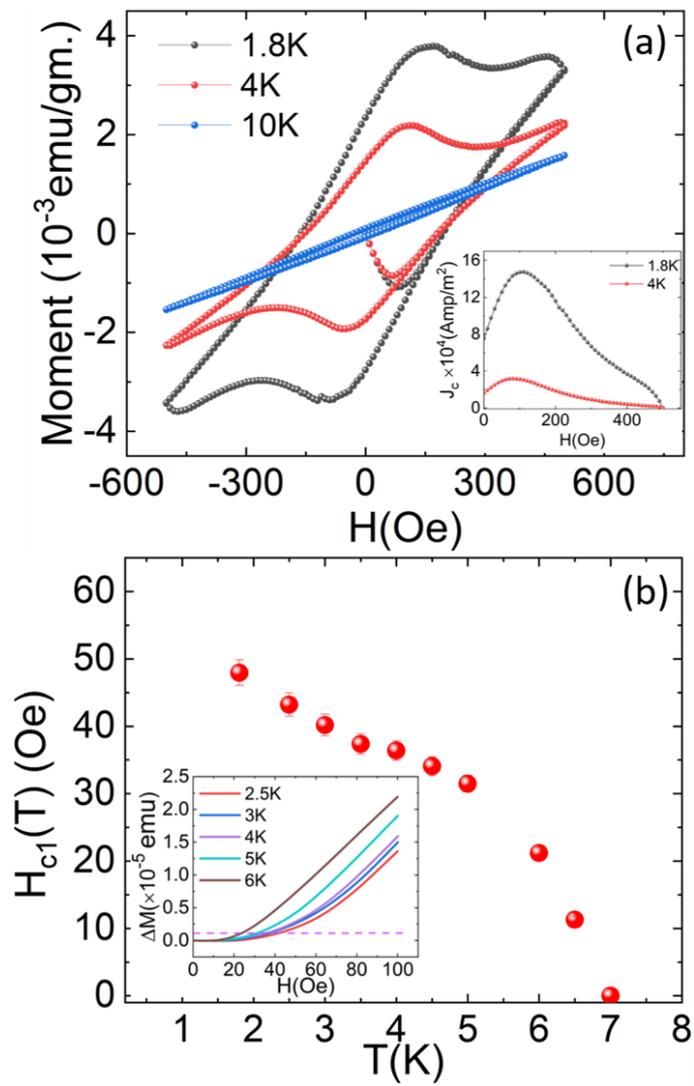

**Fig. 5**

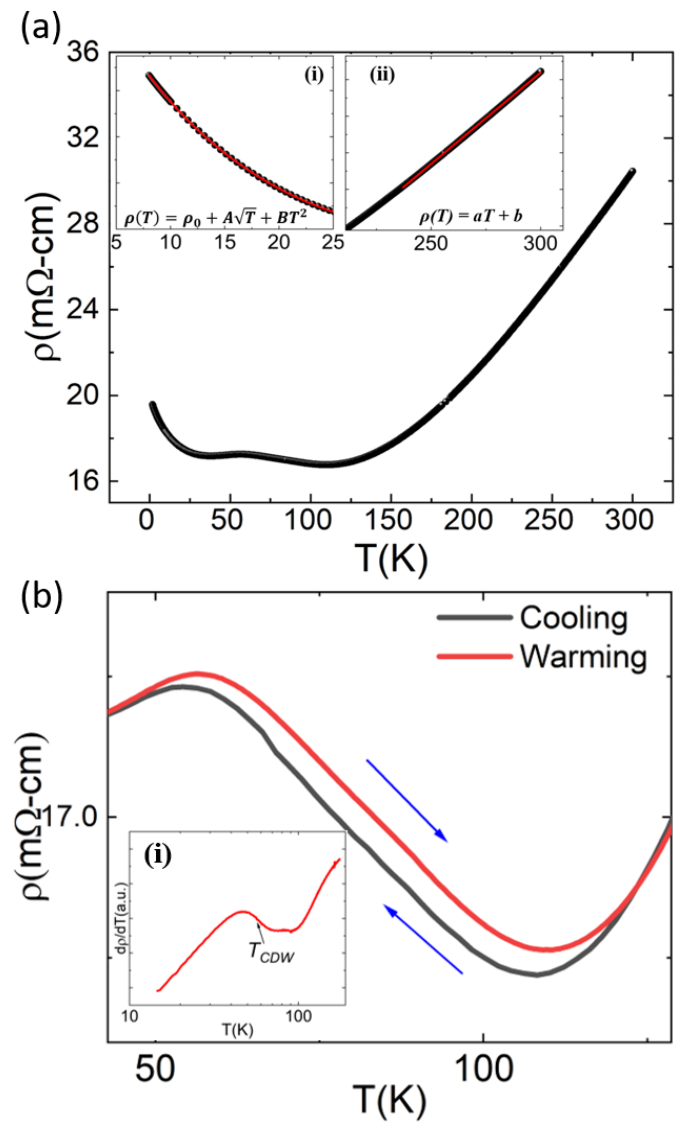

**Fig. 6**

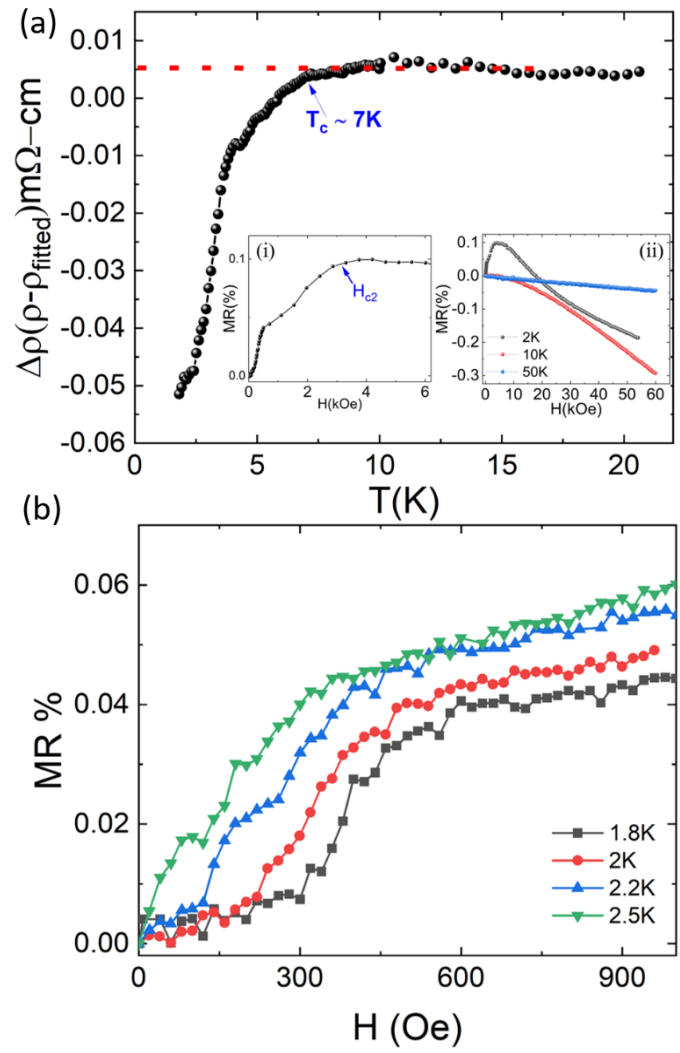